\documentclass[12pt,fleqn]{article}
\usepackage{amssymb,graphicx,epsfig}
\usepackage{lscape,graphics,amsmath,geometry}
\usepackage{latexsym}

\begin{document}
\begin{center}
\vspace*{0.7cm}

{\Large\sc {\bf Interference between $W^{\prime}$ 
and $W$ in single-top quark production processes}}
\vspace*{0.7cm}

{\sc E. Boos}, {\sc V. Bunichev},  
{\sc L. Dudko}, and {\sc  M. Perfilov}
\begin{small}
\vspace*{0.9cm}

  Skobeltsyn Institute of Nuclear Physics, Moscow State University,
119992 Moscow, Russia. \\
\vspace{3mm}
\end{small}
\end{center}
\vspace*{0.7cm}
\begin{center}
\subsubsection*{Abstract.}
\end{center}
The paper is devoted to studies of prospects to search for 
hypothetical $W^{\prime}$ gauge bosons 
at hadron colliders via single-top quark production process. 
A special attention is paid to the interference between the Standard Model 
(SM) $W$ and $W^{\prime}$ 
boson contributions. A model independent analysis is performed  
for a wide interval of $W^{\prime}$ 
masses potentially acceptable for a detection at the Tevatron and LHC.
It is shown that the interference contribution to the cross section of the most
promising s-channel single top production mode could be as large 
as 30 \% for certain
parameter points which is comparable to NLO effects computed 
in previous studies separately 
for the $W^{\prime}$ signal and the SM single-top background. 
The interference contribution affects particle distributions and 
has to be taken into account for more accurate $W^{\prime}$ signal 
and background simulation. 
\vspace*{0.7cm}

\section{Introduction}
The interactions of charged weak currents are realized in SM via exchange of 
charged massive gauge boson fields $W^+$ and $W^-$.
Although any additional charged massive bosons are not found yet
experimentally their existence is predicted by various extensions
of SM.
The wide-common name for this vector boson field is $W^{\prime}$.
Such models like Non-Commuting Extended Technicolor
~\cite{Chivukula:1995gu},
Composite~\cite{Kaplan:1983fs, Georgi:1984af} and Little higgs
models~\cite{Arkani:2001nc, Kaplan:2003uc, Schmaltz:2005ky}, models of composite gauge
bosons~\cite{composites}, Supersymmetric top-flavor models~\cite{Batra:2003nj}, Grand Unification~\cite{guts} and
Superstring theories~\cite{Cvetic:1996mf, Pati:2006nw, superstrings}
 represent examples where extension of gauge group lead to
appearing of $W^{\prime}$.
Physical properties and interaction parameters of $W^{\prime}$ can
vary from model to model. For example the $W^{\prime}$ is presented
in models of Universal extra-dimensions~\cite{Datta:2000gm,
extradimensions} as the lowest Kaluza-Klein mode of charged gauge
boson $W^{\pm}$ and has the same (V-A) chiral structure of
interaction to fermion fields as the SM $W^{\pm}$. But in top-flavor models
 the $W^{\prime}$ boson interacts differently with fermions 
of the first two and third generations. It depends on magnitude of
gauge coupling parameters $g_h$  and $g_l$. 
If $g_h > g_l$ the $W^{\prime}$ 
couples stronger to the third generation and weaker to the first two
generations, and vice versa if $g_h < g_l$~\cite{Malkawi:1996fs,Muller:1996dj}.
Frequently $W^{\prime}$ bosons are discussed in connection with so called 
Left-Right symmetric
models~\cite{Pati:1974yy,Mohapatra:1974hk,Mohapatra:1974gc,
Senjanovic:1975rk,Mimura:2002te,Cvetic:1983su,Langacker:1989xa}.
 The simplest extension of SM with Left-
Right symmetry is based on ${\mathrm U(1)} \times {\mathrm SU(2)}_L 
\times {\mathrm SU(2)}_R$ gauge group.
The left-handed fermions transform as doublets under $SU(2)_L$ and  
invariant under $SU(2)_R$ contrary to the
right-handed ones which transform as doublets under $SU(2)_R$ 
and invariant under $SU(2)_L$. Linear
combinations of charged gauge fields produce massive eigenstates: 
$W_1=\cos\zeta\, W_L + \sin\zeta\, W_R,~~W_2=-\sin\zeta\, 
W_L + \cos\zeta\, W_R$,
where $W_1$ is identified as  observed $W$ boson, and $W_2$ as a new 
$W^{\prime}$ boson, $\zeta$ 
is a mixing parameter between bosons for the right and left gauge groups.
Parameter $\zeta$ is constrained to a very small value ($\zeta < 10^{-3}$) 
to suppress (V+A) charged currents for light SM fermions in accord with 
experimental data~\cite{Donoghue:1982,Wolfenstein:1984}. 
In this case interactions of $W^{\prime}$ with fermions becomes almost purely right-handed.
In Left-Right symmetric models the parity is broken spontaneously which 
leads to different masses for the
$SU(2)_L$ and $SU(2)_R$ gauge bosons. There are two well-known 
variants of Left-Right models called $manifest$ and 
$pseudo-manifest$ for which the Cabibbo-Kobayashi-Maskawa matrices 
$V^L=V^R$ and $V^L=V^{R*}$ respectively.

Although the $W^{\prime}$ is not discovered yet there are experimental 
limits on its mass.
Various models with $W^{\prime}$ contain many 
parameters, and 
indirect constraints of $W^{\prime}$ are highly model dependent. 
Indirect searches for $W^{\prime}$ being extracted from leptonic 
and semileptonic 
decays and
also from cosmological and astrophysical data give very 
wide range for upper limits on $W^{\prime}$ mass varying from 
549 GeV up to 23 TeV~\cite{PDG:2006}. 

The direct limits on $W^{\prime}$ masses are based on hypothesis 
of purely right or left-handed interacting $W^{\prime}$ with 
SM-like coupling constants.
The limits from direct searches in leptonic decay modes  
of $W^{\prime}$ depend on a mass of a hypothetical right-handed 
neutrino. In case $M_{W^{\prime}} < M_{\nu_R}$ the decay mode
$W_R \to \nu_R,l$ is not kinematically allowed. The limits
in this case have been extracted from light jet decay mode
being $M_{W^{\prime}}>800$ GeV~\cite{D0:2004}.
In case $M_{W^{\prime}} > M_{\nu_R}$ the decay of $W^{\prime}$ 
to $\nu_R$ and $l$ is allowed, and the limit $M_{W^{\prime}}>786$ 
GeV~\cite{CDF:2001} has been obtained
in this case from the leptonic decay modes combining both
electron and muon channels. 

One of the most promising and perspective way to search
for $W^{\prime}$ is studying decays of $W^{\prime}$ to quarks of the
third generation ($W^{\prime} \to t\bar b$). This channel has relatively small
QCD background comparing to light jet channel, and it is less model 
dependent. Searches in
this channel at the Tevatron 
give a limit on $W^{\prime}$
mass $M_{W^{\prime}}>536$ GeV at 95\%~CL in assumption 
$M_{W^{\prime}}>M_{\nu_R}$. The assumption 
$M_{W^{\prime}}<M_{\nu_R}$ leads to slightly better 
limit $M_{W^{\prime}}>566$ GeV~\cite{CDF:2003} due to 
absence of the decay to ${\nu_R}$ and
correspondingly larger decay Br fraction to $t\bar b$. The 
$W^{\prime}$ boson decaying to the top and 
bottom
quarks contribute to the single-top production process.
The single-top production being interesting itself for various
aspects of the top quark physics gives perspective
channel to search for $W^{\prime}$~\cite{Simmons:1996ws}. 
In consequence that both $W$ and $W^{\prime}$ contribute to
single-top production process and have the couplings 
with the same fermion multiplets they should interfere to each 
other~\cite{Tait:2000sh}. 
It should be noted that the interference becomes possible 
only for the left-handed interacting $W^{\prime}$ component
because the SM $W$ interacts only with left-handed electroweak
currents being orthogonal to the right-handed interactions.

The aim of this paper is to study in more detail the 
interference phenomena between the SM $W$ and the $W^{\prime}$ bosons
in the single-top production process at the 
Tevatron and LHC energies.

Our paper is organized as follows. In the 2-nd section we 
present a simple analytical formula and a short analysis 
of the interference contribution. 
The 3-rd section describes results of 
numerical calculations. In the last section the summary and conclusion
are given.

\section{Interference between $W^{\prime}$ and $W$}

To consider the $W^{\prime}$ production in a model 
independent way we write down the lowest dimension effective Lagrangian 
of $W^{\prime}$ interactions
to quarks in most general form (possible higher dimension effective
operators are not taken into account in our analysis):
\begin{eqnarray}
{\cal L} = \frac{V_{q_iq_j}}{2\sqrt{2}} g_w \overline{q}_i\gamma_\mu 
\bigl( a^R_{q_iq_j} (1+{\gamma}^5) + a^L_{q_iq_j}
(1-{\gamma}^5)\bigr) W^{\prime} q_j 
+ \mathrm{H.c.} \,, 
\end{eqnarray}
where $a^R_{q_iq_j}, a^L_{q_iq_j}$ - left and right couplings of $W^\prime$ 
to quarks,
$g_w = e/(s_w)$ is the SM weak coupling constant and 
$V_{q_iq_j}$ is the SM CKM matrix element.
The notations are taken such that for so-called 
SM-like $W^{\prime}$ $a^L_{q_iq_j}=1$ and 
$a^R_{q_iq_j}=0$.

As was mentioned a promising way to search for $W^{\prime}$
is the single-top quark production processes.
There are three kinematically-different single-top production 
channels at hadron colliders, namely, $s$-channel, $t$-channel and 
associated $Wt$ channel. However for large $W^{\prime}$ mass
region, which we are interested in, the $W^{\prime}$ contribution
to the $t$- and associated $Wt$ channels becomes too small
to be detectable, and the only $s$-channel where $W^{\prime}$
may contribute as a resonance remains important. 

For the leading $s$-channel subprocess $u\bar{d} \to t\bar{b}$ 
the matrix element squared has the following form: 

\begin{eqnarray}
|M|^2 = V_{tb}^2 V_{ud}^2 (g_W)^4 \left[
\frac{(p_up_b)(p_dp_t)}{(\hat{s}-m_W^2)^2+\gamma_W^2 m_W^2} + \right.
\\ \nonumber +
2 a^L_{ud} a^L_{tb} (p_up_b)(p_dp_t)\frac{(\hat{s}-m_W^2)(\hat{s}-M_{W^\prime}^2)
+\gamma_W^2\Gamma_{W^\prime}^2}
{((\hat{s}-m_W^2)^2+\gamma_W^2 m_W^2)((\hat{s}-M_{W^\prime}^2)^2
+\Gamma_{W^\prime}^2 M_{W^\prime}^2)} +  \\ \nonumber \left.
+ \frac{( {a^L_{ud}}^2 {a^L_{tb}}^2 
+ {a^R_{ud}}^2 {a^R_{tb}}^2)(p_up_b)(p_dp_t) 
+ ({a^L_{ud}}^2 {a^R_{tb}}^2 + {a^R_{ud}}^2 {a^L_{tb}}^2)(p_up_t)(p_dp_b)}
{(\hat{s}-M_{W^\prime}^2)^2+\Gamma_{W^\prime}^2 M_{W^\prime}^2}
\right]
\end{eqnarray}
where $a^L_{ud}, a^R_{ud}$ - left and right couplings of $W^\prime$ to $u,d$ quarks,
and  $a^L_{tb}, a^R_{tb}$ - left and right couplings of $W^\prime$ to $t,b$ 
quarks.

One can rewrite the formula in terms of Mandelstam variables
using $(p_up_b) = -\hat{t}/2$ , $(p_dp_t) = (M_t^2 - \hat{t})/2$,
$(p_dp_b) = -\hat{u}/2$ , $(p_up_t) = (M_t^2 - \hat{u})/2$.
In these notations the formula reads as follows
\begin{eqnarray}
\label{formula_3}
|M|^2 = V_{tb}^2 V_{ud}^2 (g_W)^4 \left[
\frac{\hat{t} (\hat{t}-M_t^2)}{(\hat{s}-m_W^2)^2+\gamma_W^2 m_W^2} + 
\right.
\\ \nonumber +
2 a^L_{ud} a^L_{tb} \hat{t} (\hat{t}-M_t^2) \frac{(\hat{s}-m_W^2)(\hat{s}-M_{W^\prime}^2)
+\gamma_W^2\Gamma_{W^\prime}^2}
{((\hat{s}-m_W^2)^2+\gamma_W^2 m_W^2)((\hat{s}-M_{W^\prime}^2)^2
+\Gamma_{W^\prime}^2 M_{W^\prime}^2)} +  \\ \nonumber \left.
+ \frac{({a^L_{ud}}^2 {a^L_{tb}}^2 + {a^R_{ud}}^2 
{a^R_{tb}}^2)\hat{t} (\hat{t}-M_t^2) 
+ ({a^L_{ud}}^2 {a^R_{tb}}^2 + {a^R_{ud}}^2 
{a^L_{tb}}^2)\hat{u} (\hat{u}-M_t^2)}
{(\hat{s}-M_{W^\prime}^2)^2+\Gamma_{W^\prime}^2 M_{W^\prime}^2}
\right]
\end{eqnarray}
in complete agreement for the SM part (first term) with~\cite{Harris:2002md} and for
$W^\prime$ part (last term) with the result from the paper~\cite{Sullivan:2002jt}.  

The case of the SM-like $W^\prime$ corresponds to the couplings
$a^L_{ud}=a^L_{tb}=1$ and $a^R_{ud}=a^R_{tb}=0$.

One should stress that the interference (middle) term is proportional
to the left couplings only because the SM W-boson has the only
left $(V-A)$ type of the interaction. The term containing the
product of widths $\gamma_W^2\Gamma_{W^\prime}^2$ is completely negligible
comparing to ($\hat{s}-m_W^2)(\hat{s}-M_{W^\prime}^2)$ for any values
of $\hat{s}$ somewhere in the region between the $W$ and $W^\prime$
boson masses, and it makes the interference term very small
if $\hat{s}$ is very close (equal) to one of the masses.
However, in general, the interference term is not small being negative in the 
region of $M_W^2 < \hat{s} < M_{W^\prime}^2$ and positive for $\hat{s} > M_{W^\prime}^2$ 
(if the constants $a_L$ are positive). The interference term depends 
very weakly on the total width $\Gamma_{W^\prime}$ of $W^\prime$ boson.
As shown in the next section the interference 
contribution could be rather large and should be taken
into account for a signal and background simulations in searches
for $W^\prime$  if the $W^\prime$ has left-handed component in its
interaction to fermions. 

Note that we assume here the couplings $a^L_{ud}, 
a^R_{ud}, a^L_{tb}, a^R_{tb}$ to be real.
Generalization of the formula to the complex couplings is straightforward
and not given here.

After the integration over the $\hat{t}$-variable the partonic cross 
section takes the form:
\begin{eqnarray}
\label{formula_sigma}
\hat{\sigma}(\hat{s}) = \frac{\pi \alpha_W^2}{6}V_{tb}^2 V_{ud}^2 
\frac{(\hat{s}-M_t^2)^2 (2 \hat{s}+M_t^2)}{\hat{s}^2} 
\left[
\frac{1}{(\hat{s}-m_W^2)^2+\gamma_W^2 m_W^2} + \right.
\\ \nonumber +
2 a^L_{ud} a^L_{tb} \frac{(\hat{s}-m_W^2)(\hat{s}-M_{W^\prime}^2)
+\gamma_W^2\Gamma_{W^\prime}^2}
{((\hat{s}-m_W^2)^2+\gamma_W^2 m_W^2)((\hat{s}-M_{W^\prime}^2)^2
+\Gamma_{W^\prime}^2 M_{W^\prime}^2)} +  \\ \nonumber \left.
+ \frac{({a^L_{ud}}^2 {a^L_{tb}}^2 + 
{a^R_{ud}}^2 {a^R_{tb}}^2 + {a^L_{ud}}^2 {a^R_{tb}}^2 
+ {a^R_{ud}}^2 {a^L_{tb}}^2)}
{(\hat{s}-M_{W^\prime}^2)^2+\Gamma_{W^\prime}^2 M_{W^\prime}^2}
\right]
\end{eqnarray}
where $\alpha_W = g_W^2/(4\pi)$ and $\hat{s}=x_u x_d s$.
The well-known SM cross section (first term) completely agrees 
with~\cite{Smith:1996ij}. 

\section{Numerical illustrations}
Numerical computations and Monte Carlo simulations have been performed for the Tevatron and LHC 
energies $\sqrt{s}=1.96$ TeV and 14 TeV. 
Top quark mass was chosen $M_t=175$ GeV. Results are given for six different sets of
$W^{\prime}$ masses from 0.5 up to 1 TeV (for Tevatron) and from 0.6 up to 5 TeV (for LHC) 
separately for both pure right and left-handed interacting $W^{\prime}$. Partonic
distribution  functions CTEQ6l1 have been used.  The QCD scale has been
set to $M_{W^{\prime}}$. The couplings of $W^{\prime}$ to SM-fields 
have been implemented into CompHEP~\cite{Pukhov:1999gg} 
which was used to compute the $W^{\prime}$ width,  (see the Table~\ref{tab_width}),
production cross sections, kinematical distributions
and to generate unweighted events for different sets of $W^{\prime}$ masses.
For the case of left-handed interacting $W^{\prime}$ we set  
$a^L_{ud}=a^L_{tb}=1,~a^R_{ud}=a^R_{tb}=0$, for the right-handed case 
$a^L_{ud}=a^L_{tb}=0,~a^R_{ud}=a^R_{tb}=1$ and
for SM all the parameters are equal to zero $a^L_{ud}=a^L_{tb}=a^R_{ud}=a^R_{tb}=0$.

We simulate the process $p\bar p~(pp) \to W/W^{\prime} \to t\bar b$ 
which includes 8 subprocesses with different parton
combinations in the initial states.
\begin{table}
\begin{center}
\begin{tabular}{|c|c|c|c|c|c|c|} \hline
$M_{W^{\prime}}~~[GeV]$ & $500$ & $600$ & $700$ & $800$ & $900$ & $1000$\\
\hline
$\Gamma_{W^\prime}~~[GeV]$ & $16.14$ & $19.65$ & $23.12$ & $26.58$ & $30.01$ & $33.44$\\
\hline
\end{tabular}
\end{center}
\caption{\label{tab_width} The total width of the $W^{\prime}$ in dependence
  on $W^{\prime}$ mass for the top quark mass$M_t=175$ GeV assuming that
decays to both quarks and leptons are allowed (if $W^{\prime}$ decays to leptons
are not allowed the widths will be smaller by a factor about 3/4)}
\end{table}
\begin{table}
\begin{center}
\begin{tabular}{|c|c|c|c|c|c|}
\hline
Mass $W^\prime$, GeV & \multicolumn{3}{|c|}{SM+left $W^\prime$,} & \multicolumn{2}{|c|}{SM+right $W^\prime$,}\\
&${\sigma}_{tot}$,[pb]& IT, \% &${\sigma}_{u\bar{d} \to t\bar{b}}$ , \% & ${\sigma}_{tot}$,[pb] & ${\sigma}_{u\bar{d} \to t\bar{b}}$ , \% \\
\hline
500 & 2.13 & 12.4 & 99.0 & 2.39 & 98.8 \\
\hline
600 & 0.846 & 21.2 & 98.9 & 1.02 & 98.7 \\
\hline
700 & 0.403 & 30.8 & 98.2 & 0.524 & 98.1 \\
\hline
800 & 0.256 & 33.4 & 97.3 & 0.341 & 97.4 \\
\hline
900 & 0.212 & 30.5 & 96.6 & 0.275 & 96.8 \\
\hline
1000 & 0.202 & 23.5 & 96.4 & 0.25 & 96.6\\
\hline
\end{tabular}
\end{center}
\caption{\label{tab_tev_cs} Total cross section of the process
         $p\bar p \to W/W^{\prime} \to t\bar b $;
          contribution of the interference term (IT, in \%) and contribution of leading 
         $u\bar{d} \to W/W^{\prime} \to t\bar{b}$ subprocess to the total cross section
	 for various $W^{\prime}$ masses and Tevatron energy ($\sqrt{s}=1960$ GeV). Interference term is zero for the right interacting $W^{\prime}$.}
\end{table}

\begin{table}
\begin{center}
\begin{tabular}{|c|c|c|c|c|c|} \hline
$M_{W^{\prime}}$, [GeV] & \multicolumn{3}{|c|}{SM+left $W^\prime$} & \multicolumn{2}{|c|}{SM+right $W^\prime$}\\
 & ${\sigma}_{tot}$, [pb] & IT, \% &${\sigma}_{u\bar{d}(\bar{d}u) \to t\bar{b}}$ , \%  & ${\sigma}_{tot}$, [pb] & ${\sigma}_{u\bar{d}(\bar{d}u) \to t\bar{b}}$ , \% \\
\hline
600 & 37.3 & 8.65 & 90.3 & 40.6 & 90.0 \\
\hline
800 & 16.1 & 12.6 & 90.9 & 18.2 & 90.5 \\
\hline
1000 & 9.42 & 14.1 & 90.2 & 10.7 & 91.0 \\
\hline
5000 & 4.91 & 1.00 & 85.5 & 4.96 & 85.5 \\
\hline
\end{tabular}
\end{center}
\caption {\label{tab_lhc_cs} 
Total cross section of the process
         $pp \to W/W^{\prime} \to t\bar b $,
          contribution of the interference term (IT, in \%) and contribution of leading 
         $u\bar{d}(\bar{d}u) \to W/W^{\prime} \to t\bar{b}$ subprocess to the total cross section
	 for various $W^{\prime}$ masses and LHC energy ($\sqrt{s}=14$ TeV). Interference term is zero for the right interacting $W^{\prime}$.
 }
\end{table}

In the Tables~\ref{tab_tev_cs},~\ref{tab_lhc_cs} 
the total cross section of the process $p\bar p~(pp) \to W/W^{\prime} \to t\bar b$,
the contribution of the interference term (in \%) and 
the contribution of the leading
subprocess $u\bar{d} \to W/W^{\prime} \to t\bar{b}$ to the
total cross section are listed as a function of $W^{\prime}$ masses.
The tables show that the contribution from subleading subprocesses is 
small for all values of $W^{\prime}$ mass. 
The cross sections for the purely left-handed
interacting $W^{\prime}$ are smaller than for the right-handed one. 
This difference reflects the fact of negative
contribution of the interference between the left-handed interacting $W^{\prime}$ 
and the SM $W$-boson which absent in the right-handed case.
One can see that for the Tevatron (Table~\ref{tab_tev_cs}) the interference 
term achieves a maximum about 33,4 \% of the total cross
section at $M_{W^{\prime}}$ equal to 800 GeV, and about 14 \% for the
LHC (Table~\ref{tab_lhc_cs}) at 1 TeV $W^{\prime}$ boson mass.
\begin{figure*}
\begin{minipage}[b]{.49\linewidth}
\centering
\includegraphics[width=70mm,height=60mm]{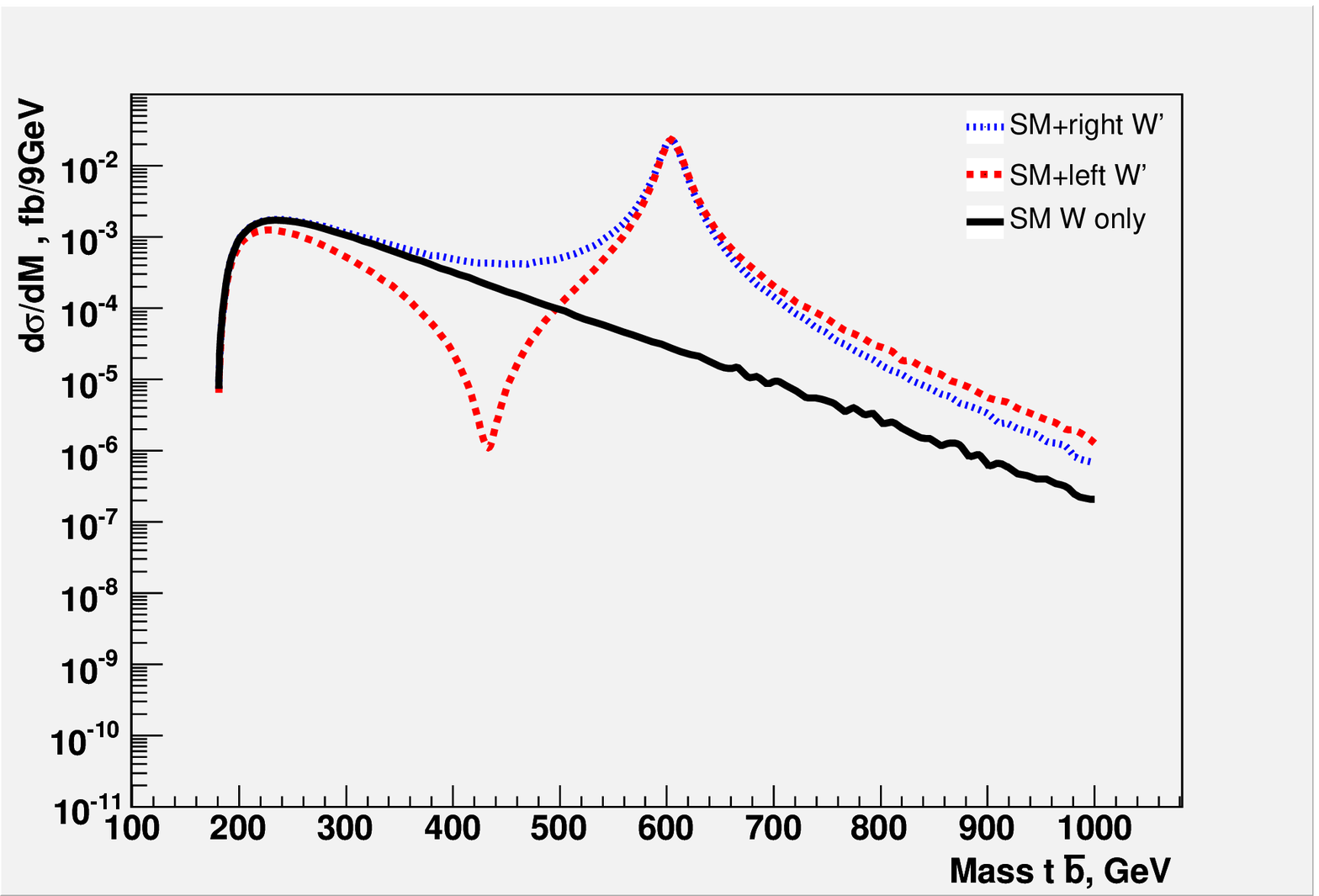}
\includegraphics[width=70mm,height=60mm]{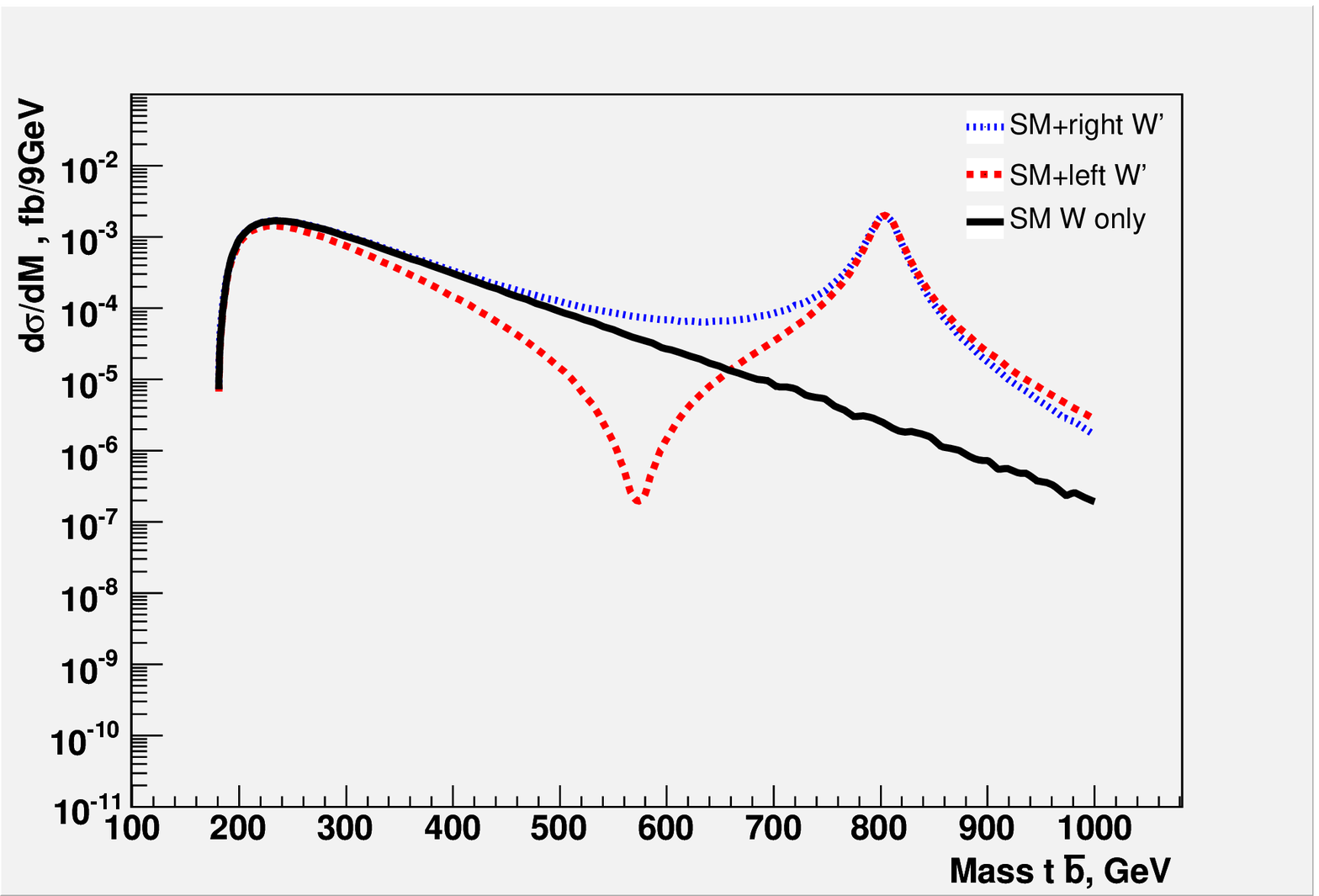}
\includegraphics[width=70mm,height=60mm]{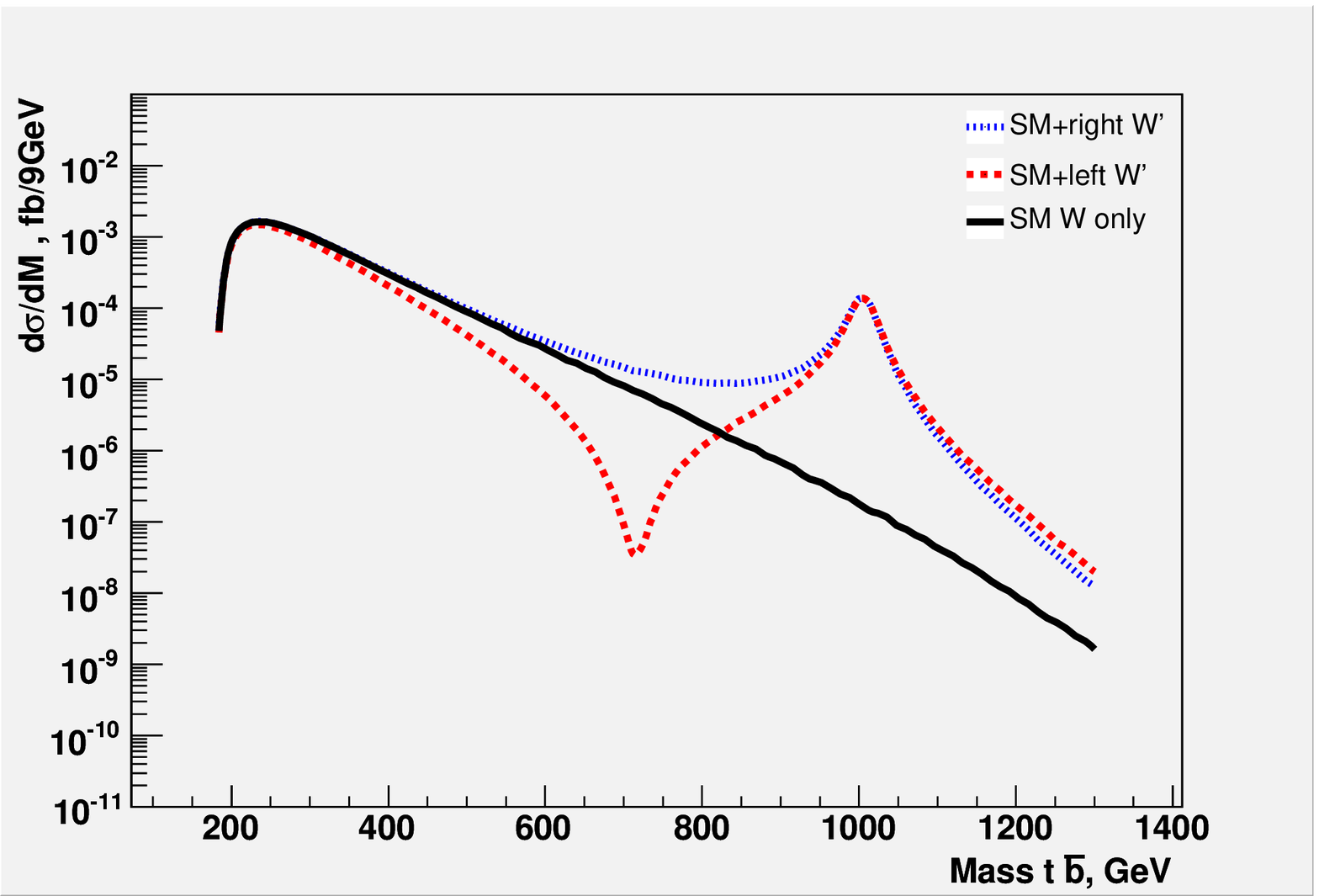}
\caption{Invariant mass of $t\bar{b}$ system for $M_{W^{\prime}}$ equal 600 GeV, 800 GeV and 1000 GeV respectively at the Tevatron.} \label{inv_mass_tev}
\end{minipage}
\begin{minipage}[b]{.49\linewidth}
\centering
\includegraphics[width=70mm,height=60mm]{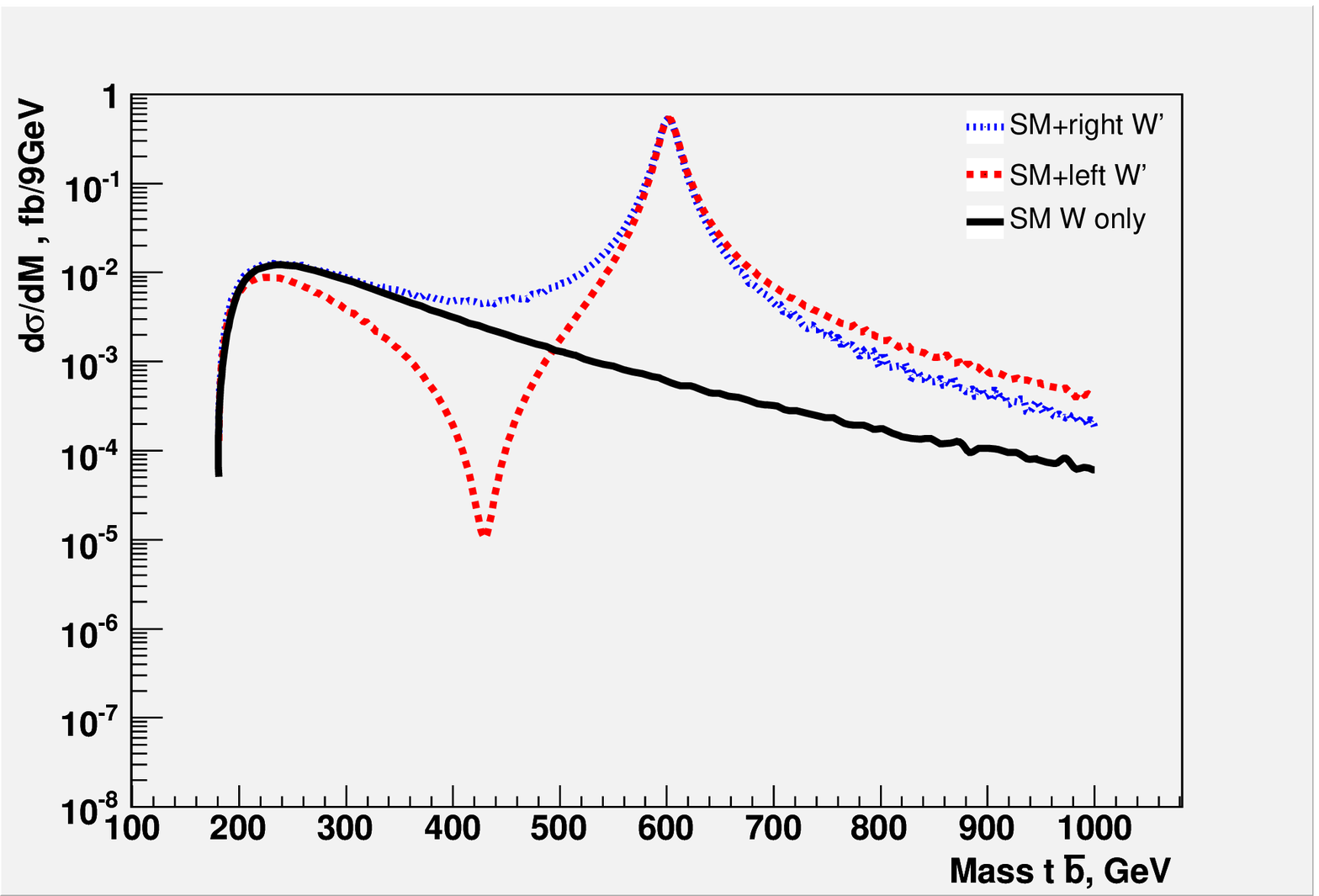}
\includegraphics[width=70mm,height=60mm]{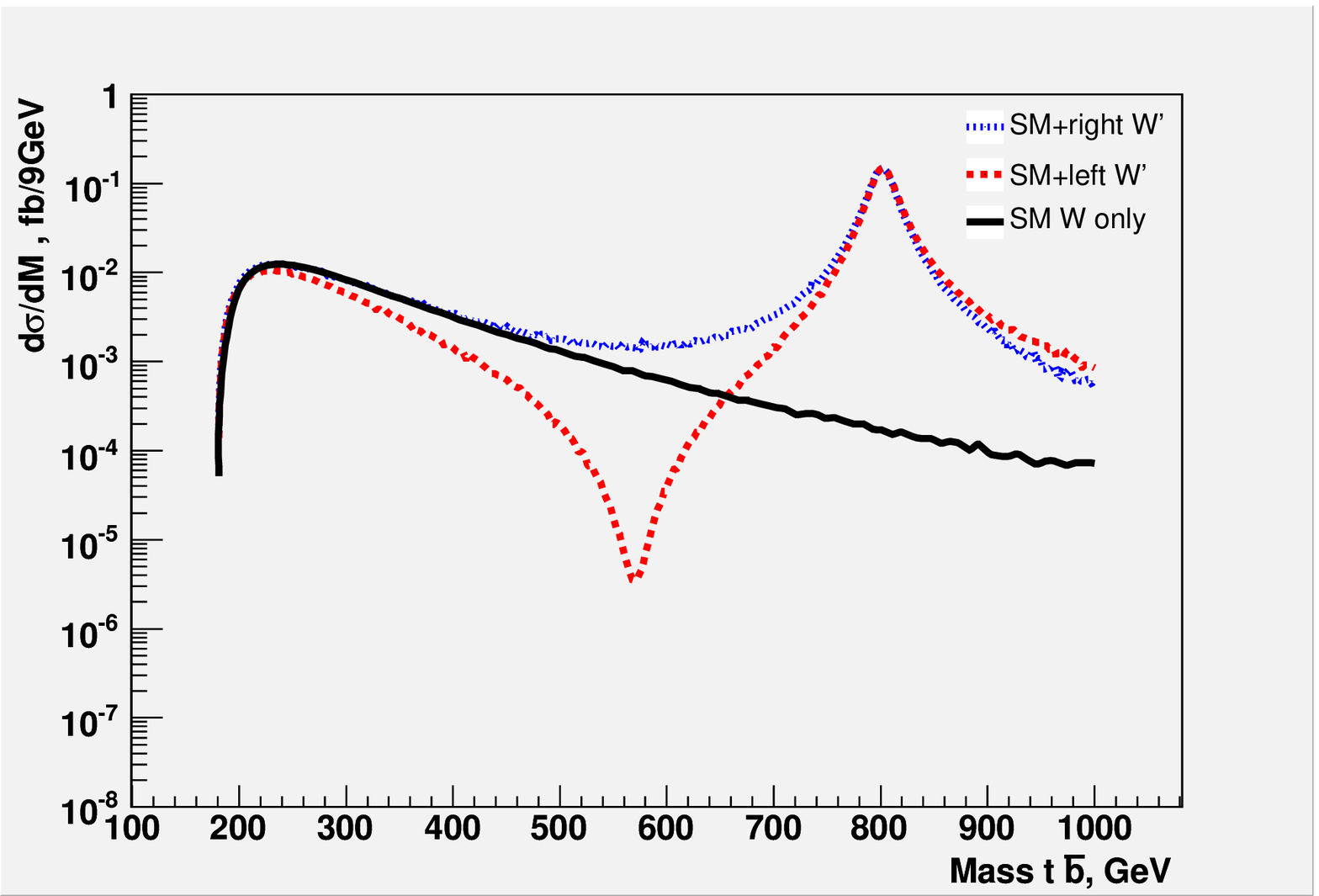}
\includegraphics[width=70mm,height=60mm]{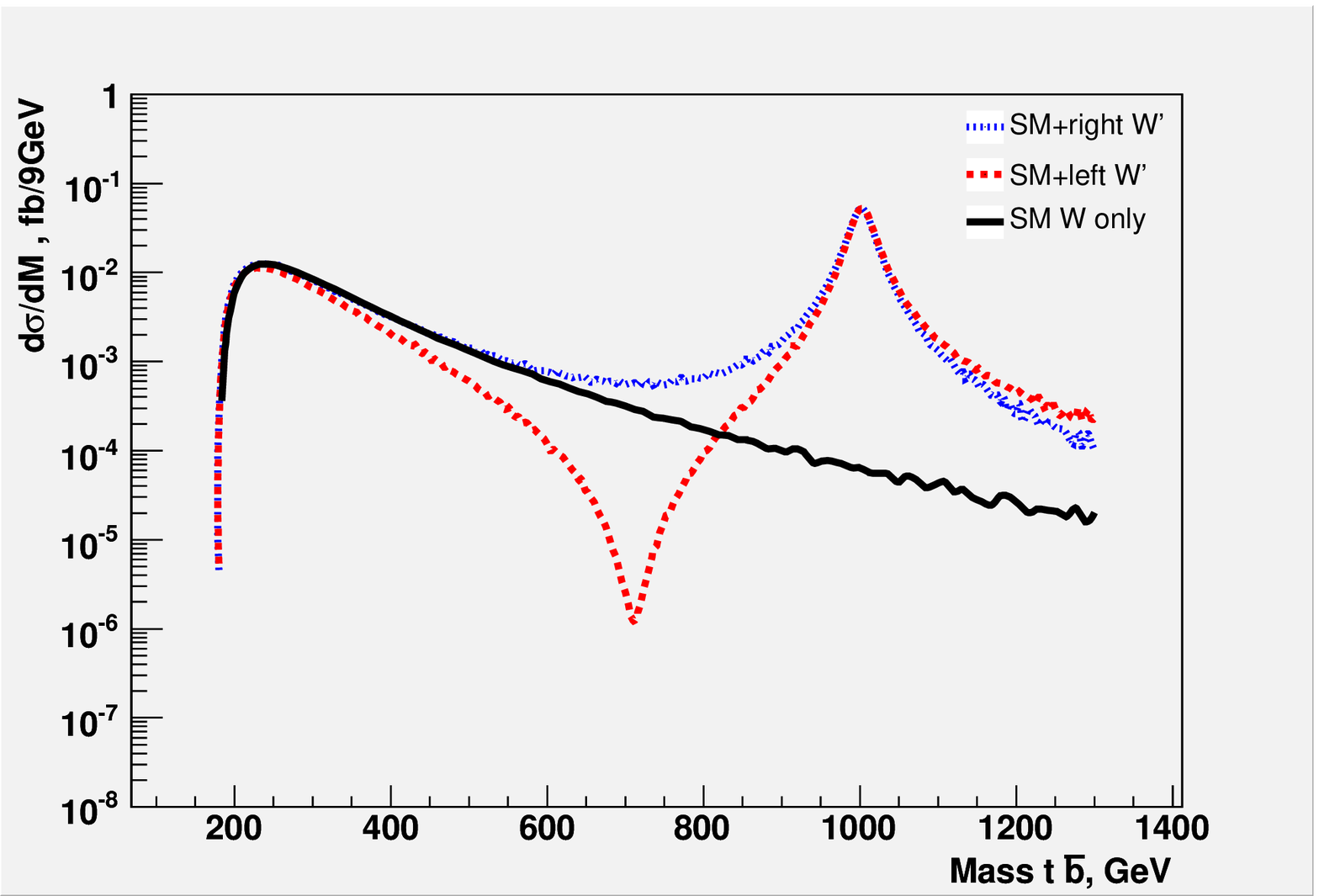}
\caption{Invariant mass of $t\bar{b}$ system for $M_{W^{\prime}}$ equal 600 GeV, 800 GeV and 1000 GeV respectively.at the LHC} \label{inv_mass_lhc}
\end{minipage}
\end{figure*}
\begin{figure*}
\begin{minipage}[t]{.49\linewidth}
\centering
\includegraphics[width=70mm,height=60mm]{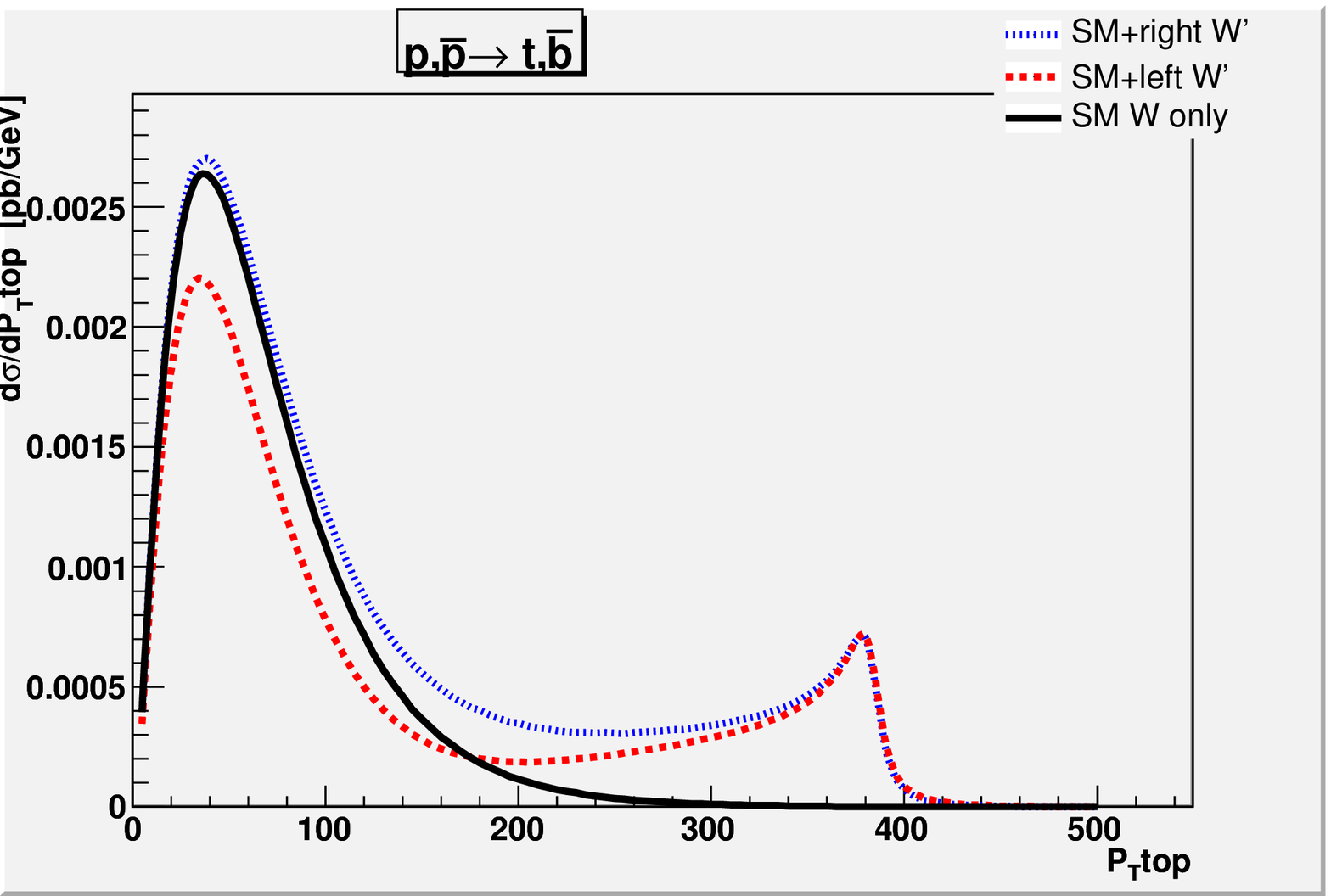} 
\caption{\label{fg:tev_ptt} Top quark transverse momentum $P_T^t$ (Tevatron)}
\end{minipage}
\begin{minipage}[t]{.49\linewidth}
\centering
\includegraphics[width=70mm,height=60mm]{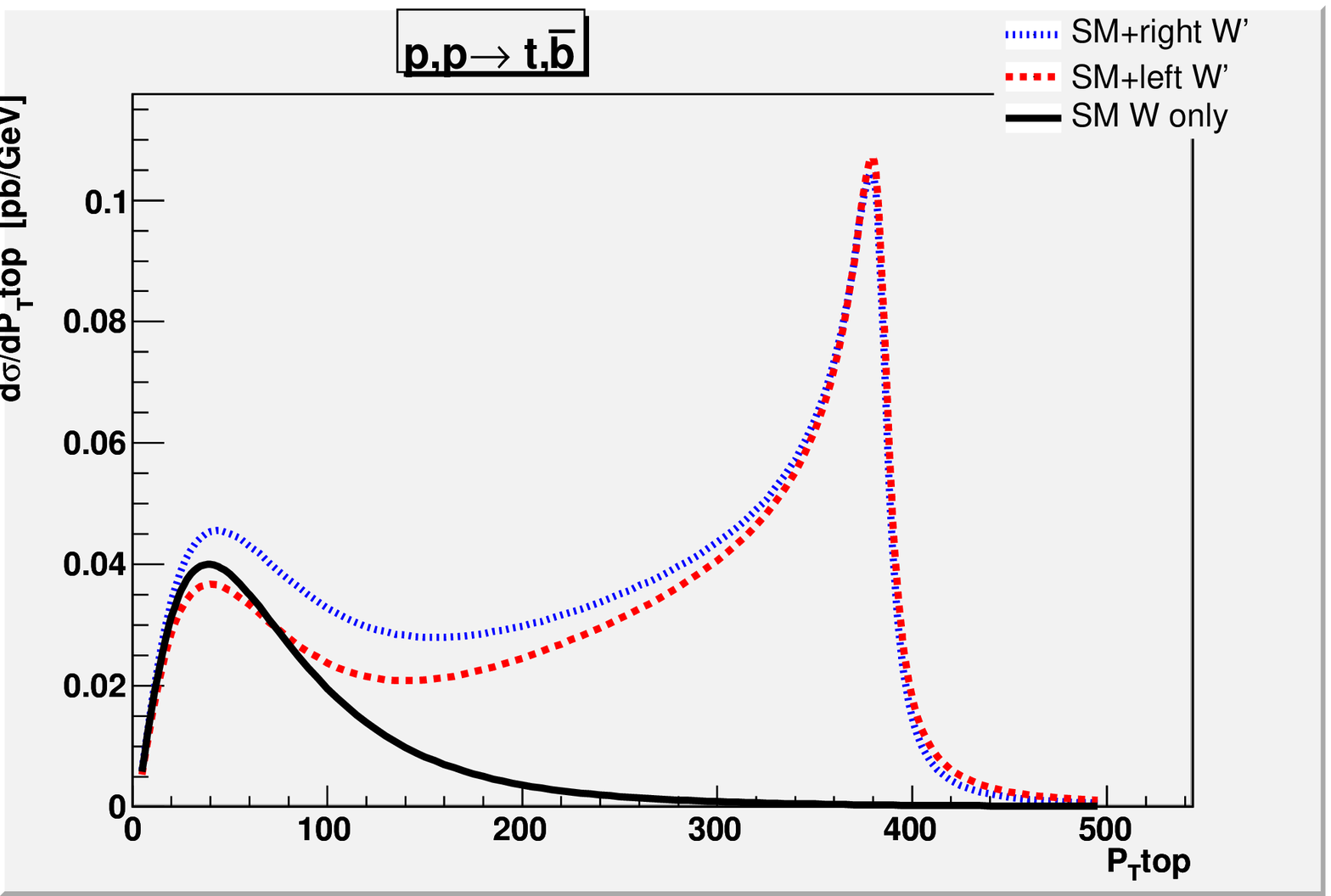}
\caption{\label{fg:lhc_ptt} Top quark transverse momentum $P_T^t$ (LHC)}
\end{minipage}
\vspace*{1cm}

\begin{minipage}[t]{.49\linewidth}
\centering
\includegraphics[width=70mm,height=60mm]{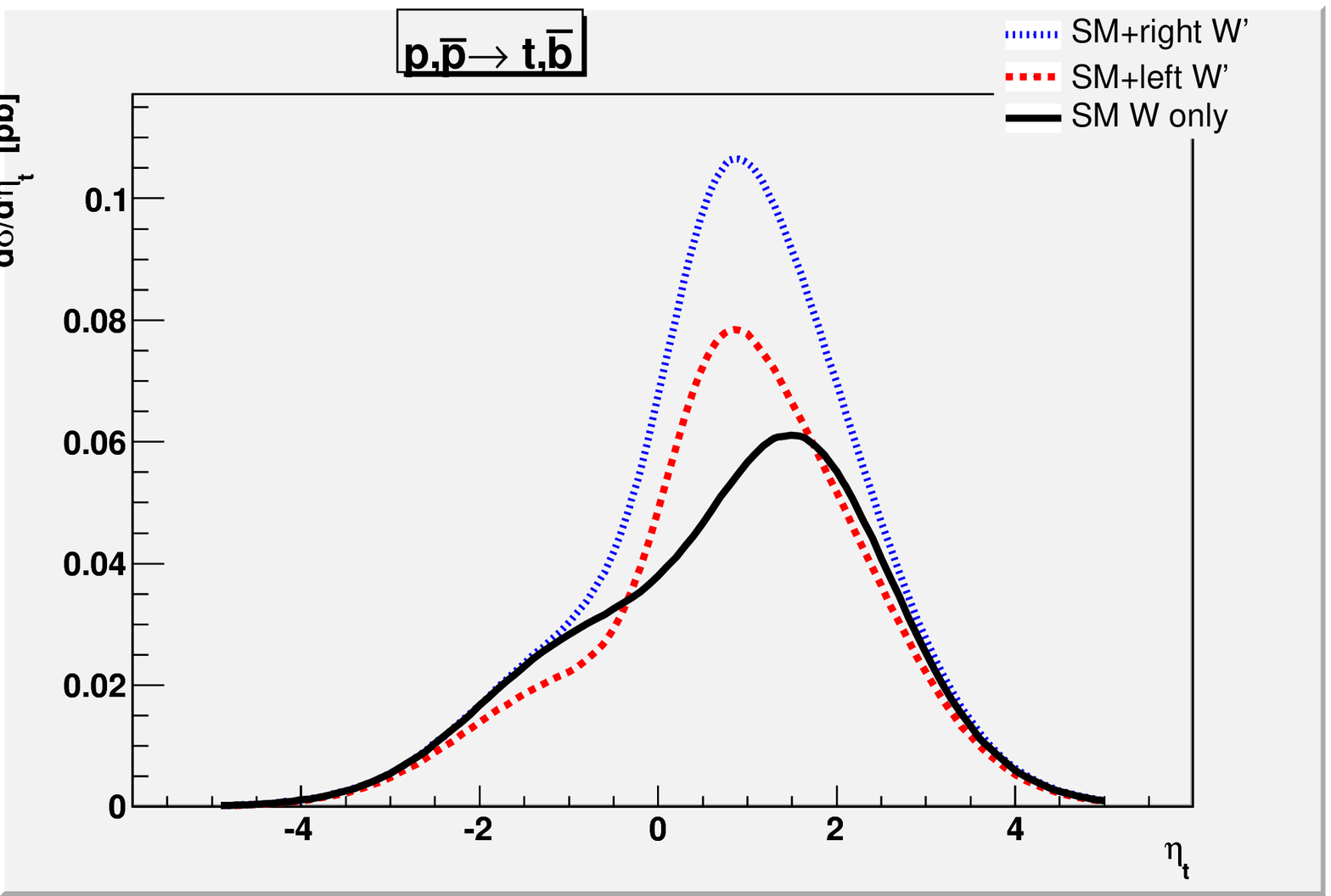} 
\caption{\label{fg:tev_etat} Top quark pseudorapidity $\eta^t$ (Tevatron)}
\end{minipage}
\begin{minipage}[t]{.49\linewidth}
\centering
\includegraphics[width=70mm,height=60mm]{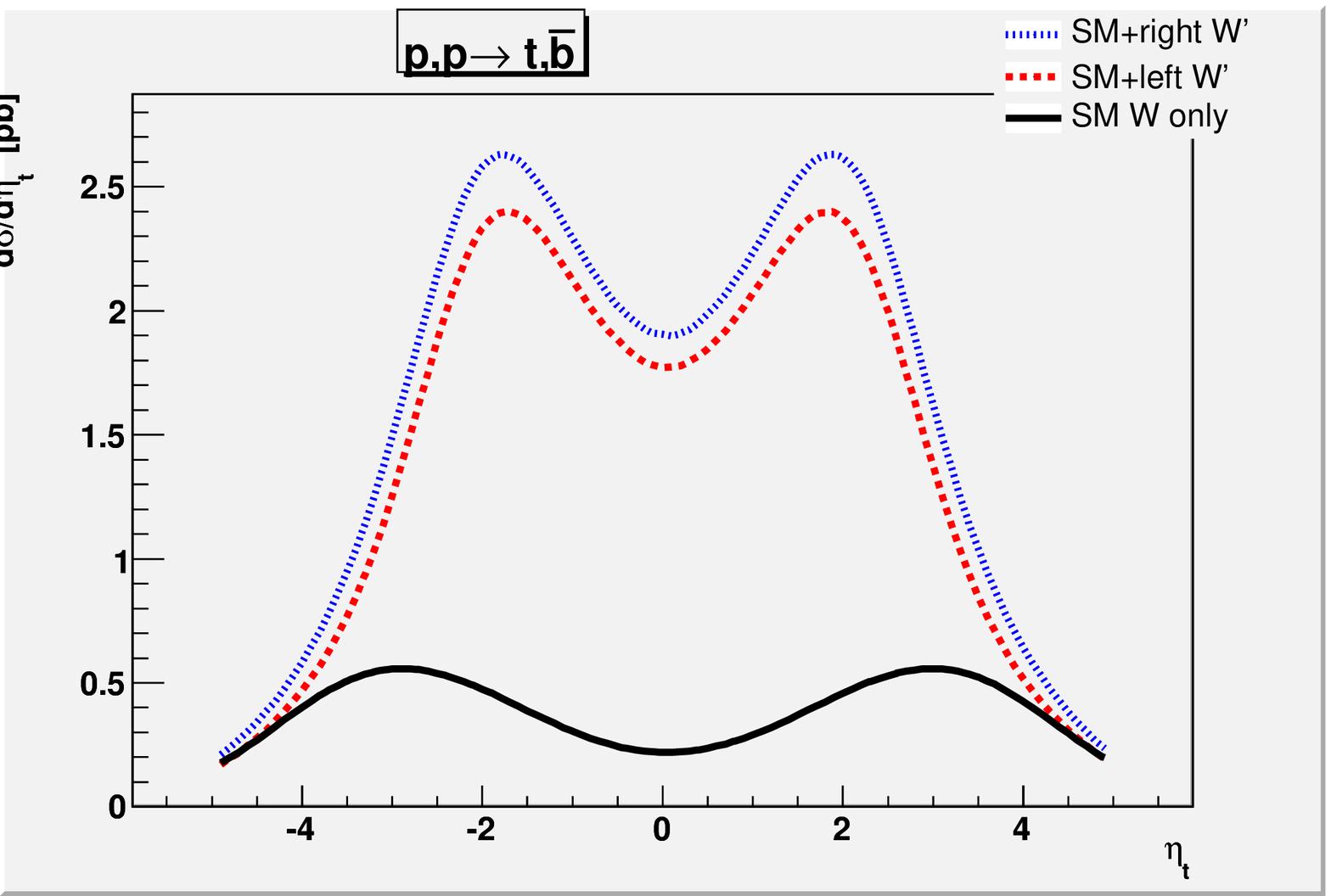}
\caption{\label{fg:lhc_etat} Top quark pseudorapidity $\eta^t$ (LHC)}
\end{minipage}
\end{figure*}

\begin{figure*}
\begin{minipage}[t]{.49\linewidth}
\centering
\includegraphics[width=70mm,height=60mm]{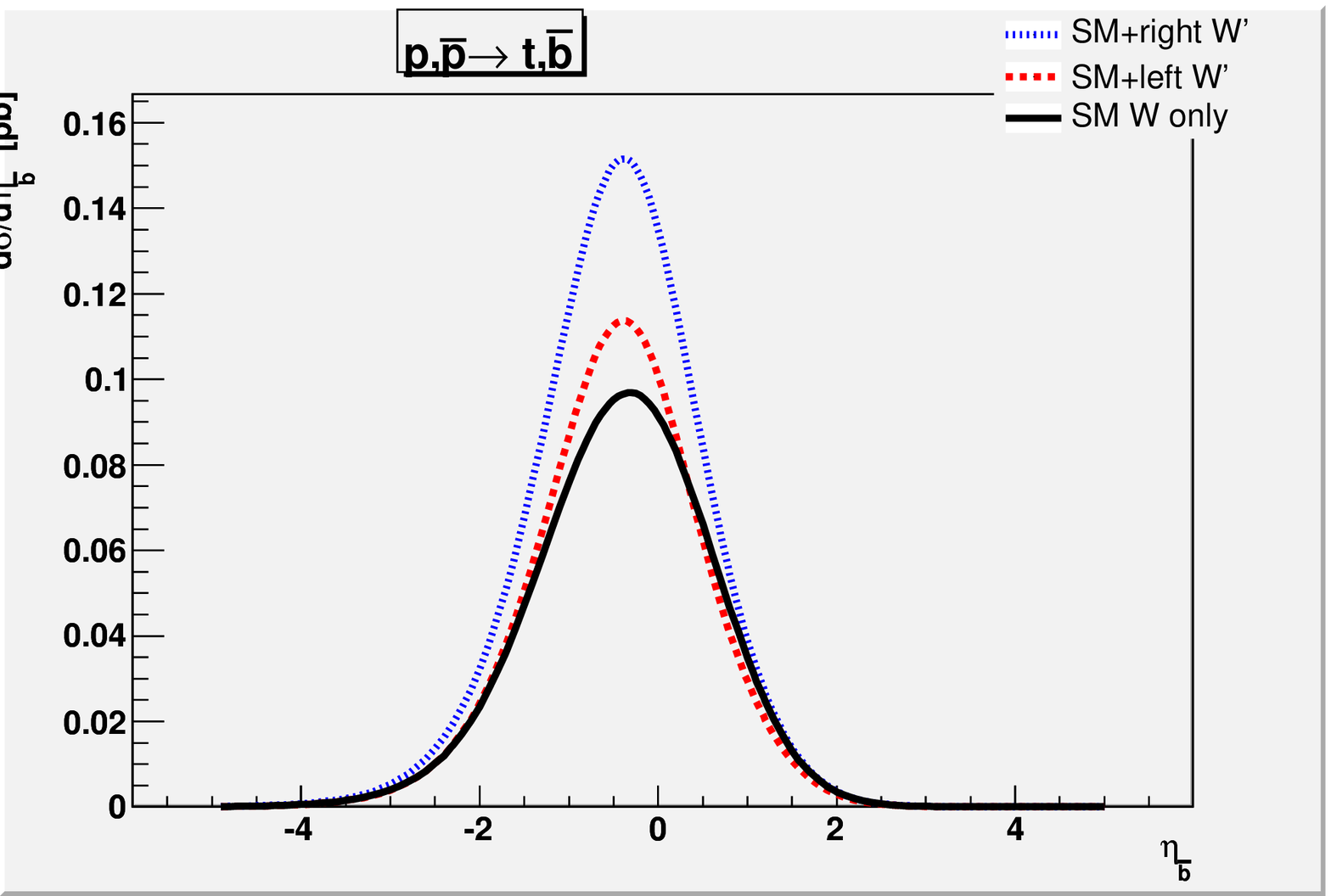} 
\caption{\label{fg:tev_etab2} $\bar{b}$ quark pseudorapidity $\eta^{\bar{b}}$ (Tevatron)}
\end{minipage}
\begin{minipage}[t]{.49\linewidth}
\centering
\includegraphics[width=70mm,height=60mm]{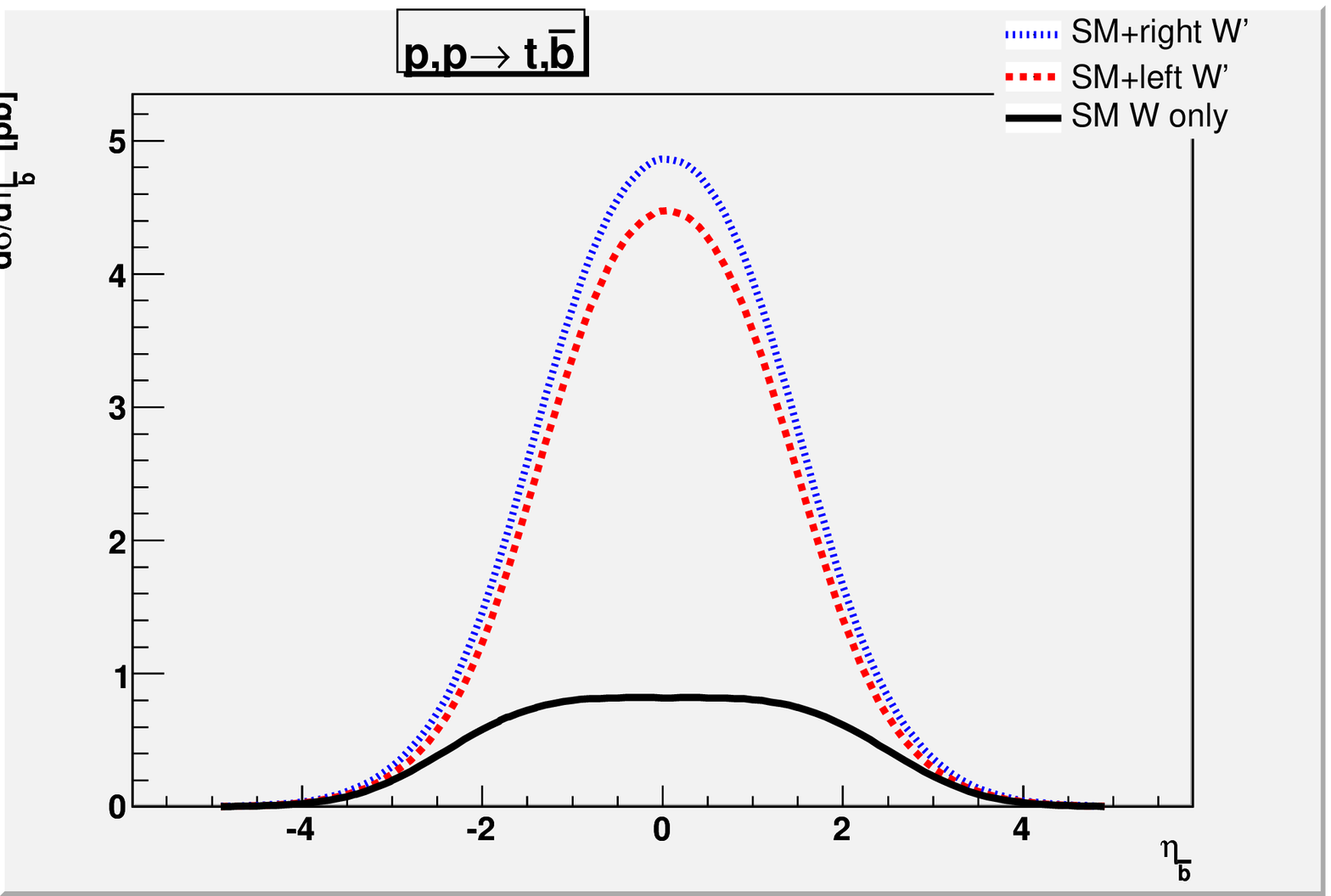}
\caption{\label{fg:lhc_etab2} $\bar{b}$ quark pseudorapidity $\eta^{\bar{b}}$ (LHC)}
\end{minipage}
\end{figure*}

To illustrate the interference between the $W^{\prime}$ and the SM $W$ 
bosons in more detail the Fig.~\ref{inv_mass_tev} and Fig.~\ref{inv_mass_lhc}
show the differential cross section for the s-channel
single-top quark production as a function of  the
invariant mass of $tb$-system for three sets of $W^{\prime}$ 
masses 600 (upper plot), 800 (middle plot), and 1000 GeV (lower plot). 
For each value of $W^{\prime}$ masses the cases
with purely left-handed and right-handed interactions of $W^{\prime}$ 
are plotted comparing to the SM process. 
All curves start from the reaction threshold 
at about $M_{t\bar{b}}\approx$ 180 GeV. 
In  case of right-handed interactions of $W^{\prime}$ there is no
interference, and  the curve for the invariant mass distribution
is the algebraic sum of two independent falling down SM $W$ and the
resonant $W^{\prime}$ distributions.
The picture is significantly different in case of left-handed 
interacting $W^{\prime}$ where in addition to the resonance pike there 
is an area with a minimum due to destructive interference between
the SM $W$ and the $W^{\prime}$ boson contributions.  
This local minimum follows form the formula~\ref{formula_sigma} and 
takes place at the value of $tb$ invariant mass 
$M_{tb}=\sqrt{\frac{M_{W^{\prime}}^2 + M_W^2}{2}}$. 

The Figs.~\ref{fg:tev_ptt}-\ref{fg:lhc_etat} 
and Figs.~\ref{fg:tev_etab2}-\ref{fg:lhc_etab2}.
are present a momentum transfer ($P_T$) and pseudorapidity ($\eta$) distributions 
of the $t$- and $\bar{b}$-quarks produced at the Tevatron and LHC.
These plots once more illustrate the points discussed above.
In case of SM + right interacting $W^{\prime}$ the plot of 
top-quark $P_T$ (Figs.~\ref{fg:tev_ptt}) 
represents an algebraic sum of two distributions coming from SM W and from $W^{\prime}$. 
The situation is changed in case of
SM~+~left interacting $W^{\prime}$, where the presence of the interference 
leads to a reduction of the cross section, and correspondingly the curve for
this case lies below the curve for the case SM~+~right interacting 
$W^{\prime}$. 
The picture for the LHC
(Figs.~\ref{fg:lhc_ptt}) differs from the Tevatron case only by larger 
relative contribution of the $W^{\prime}$ boson decaying to the top 
quark. Pseudorapidity distributions of the top quark 
in the Figs.~\ref{fg:tev_etat},\ref{fg:lhc_etat} also
demonstrate a reduction of the rate  for the case of SM~+~left 
interacting $W^{\prime}$ in
comparison to the  SM + right interacting $W^{\prime}$. 
In contrast to the LHC (Figs.~\ref{fg:lhc_etat})
the Tevatron distribution shown for the top 
quark only (Figs.~\ref{fg:tev_etat}) is asymmetric 
because of the difference in PDF for the proton and antiproton. 
The plots for the b-quark $P_T$ represent distributions
similar (equivalent for ideally reconstructed top decay products) with 
those for the top quark, but the
pseudorapidity distributions of b-quarks are 
more central (Figs.~\ref{fg:tev_etab2},\ref{fg:lhc_etab2})
comparing to the top quark.
 
We have simulated the complete chain of top quark decays
taking into account all the spin correlations between 
top quark production and its subsequent decay. The separate study
of angular correlations for these processes 
is given in the paper~\cite{wprime_spin_correlations}.

\section{Conclusion}

In this paper we focus our attention on the interference of SM $W$
gauge boson with a hypothetical $W^{\prime}$ vector boson
in single-top production process at hadron collider.
Simple symbolic formula for matrix element squared
of the leading subprocess  shows  explicitly in a model
independent way a parameter
dependence for general left- and right-handed couplings of 
$W^{\prime}$ boson to the  SM fermions. As expected
a maximal influence of the interference term takes
place for the case of left-handed interactions of $W^{\prime}$.
Such a destructive interference may approach as large
as 30\% for the Tevatron leading to a local minimum in
the invariant mass distribution at 
$M_{tb}=\sqrt{\frac{M_{W^{\prime}}^2 + M_W^2}{2}}$ and being very small close
to the resonance position in $tb$ invariant mass.   
The NLO corrections have been computed separately for
the SM single top s-channel process~\cite{Smith:1996ij} (background 
for $W^{\prime}$) and for the $W^{\prime}$ contribution~\cite{Sullivan:2002jt}. 
Such an approach works perfectly for the case of
purely right-handed interacting $W^{\prime}$. However 
a recipe how to proceed  from the right-handed to a general
coupling case by multiplying the answer by the coefficient 
${a^L_{ud}}^2 {a^L_{tb}}^2 + {a^R_{ud}}^2 {a^R_{tb}}^2 $
obviously does not work because
the interference contribution and the additional term in $W^{\prime}$
contribution are proportional to different coupling combinations,
the first to $a^L_{ud} a^L_{tb}$ and 
the second 
${a^L_{ud}}^2 {a^R_{tb}}^2 + {a^R_{ud}}^2 {a^L_{tb}}^2$ (see, Formulas~\ref{formula_3},~\ref{formula_sigma}).
As follows from the Formulas~\ref{formula_3},~\ref{formula_sigma} 
in order to simulate the general coupling dependence
one should perform the simulations for three cases,
purely right-handed ($a^R_{tb}=a^R_{ud}=1$, others equal to 0), purely
left-handed ($a^L_{tb}=a^L_{ud}=1$, others equal to 0), and mixed
(for example, $a^R_{tb}=1,a^L_{ud}=1$, others equal to 0).
Computation of the NLO corrections in general case
including non-trivial interference remains to be done, however
in a number of cases the NLO corrections can be extracted
from the existing computations.

The interference contribution is important
in general and has to be taken into account 
in searches for $W^{\prime}$ bosons performing MC event generation. 
 Calculations described in this paper 
and correspondingly created Monte-Carlo event generator 
have been used in recent experimental searches for $W^{\prime}$ 
by the D\O collaboration~\cite{Abazov:2006aj}. The special study
of angular correlations in the production and decay 
of $W^{\prime}$ is described in the separate paper~\cite{wprime_spin_correlations}.

\subsubsection*{Acknowledgments:}
We acknowledge a support of Russian Foundation for Basic Research 
(grant RFBR~04-02-17448) 
and Russian Ministry of Education and Science (grant NS.8122.2006.2).

\end{document}